\documentclass[aps,showpacs,groupedaddress]{revtex4}  % for review and submission 
\usepackage{graphicx}  % needed for figures
\usepackage{dcolumn}   % needed for some tables
\usepackage{bm}        % for math
\usepackage{amssymb}   % for math
\usepackage{amsmath}
\usepackage{epsfig}
\usepackage{xspace}
\usepackage{color}

\newcommand{\bea}{\begin{eqnarray}}
\newcommand{\eea}{\end{eqnarray}}

\begin{document}

% the following line is for submission, including submission to the arXiv!!
%\hspace{5.2in} \mbox{Darmstadt/DESY/U-Tokyo}

\title{Complex Langevin dynamics for dynamical QCD at nonzero chemical potential: a comparison with multi-parameter reweighting}
\author{Z. Fodor$^{1,2,3}$}
\author{S. D. Katz$^{3,4}$}
\author{D. Sexty$^{1}$}
\author{C. T\"or\"ok$^{3,4}$}

\affiliation{ $^1$ {\it Department of Physics, Wuppertal University, Gaussstr. 20, D-42119 Wuppertal, Germany}}
\affiliation{$^2$ {\it J\"ulich Supercomputing Centre, Forschungszentrum J\"ulich, D-52425 J\"ulich, Germany} }
\affiliation{ $^3$ {\it Inst. for Theoretical Physics, E\"otv\"os University, P\'azm\'any P. s\'et\'any 1/A, H-1117 Budapest, Hungary}}
\affiliation{ $^4$ {\it MTA-ELTE "Lend\"ulet" Lattice Gauge Theory Research Group, P\'azm\'any P. s\'et\'any 1/A, H-1117 Budapest, Hungary}}

\begin{abstract} 
We study lattice QCD at non-vanishing chemical potential using
the complex Langevin equation. We compare the results with multi-parameter
reweighting both from $\mu=0$ and phase quenched ensembles. We find a good agreement for lattice spacings below $\approx$0.15 fm. On coarser lattices the complex Langevin approach breaks down. Four flavors of staggered fermions are used on $N_t=4, 6$ and 8 lattices. For one ensemble we also use two flavors to investigate the effects of rooting. 
\end{abstract}

\maketitle

\section{Introduction and overview}\label{sec:introduction}

Dense and/or high temperature phases of strongly interacting matter 
are becoming experimentally 
accessible nowadays due to heavy ion collision experiments
at the Relativistic Heavy Ion Collider, the Large Hadron Collider, and 
especially the FAIR facility at GSI, as well as astrophysical observations 
of neutron starts. Theoretical understanding of the dense,
strongly interacting phases and 
the first principles determination of the phase diagram
of QCD as a function of the temperature and chemical potential 
are still lacking. This is a consequence of the sign problem, 
which makes lattice calculations 
at nonzero baryon density challenging. 

 The standard non-perturbative tool for QCD, lattice QCD is defined by 
the path integral
\bea 
  Z = \int DU e ^{-S_{YM}} \textrm{det} M(\mu)
\eea
with the Yang-Mills action $S_{YM}$ of the gluons and the fermion 
determinant $M(\mu)$ on a cubic space-time lattice.  
At nonzero chemical potential the determinant 
is non-real, therefore importance sampling methods are not 
applicable. For a review of ideas to circumvent the sign problem see 
 \cite{Fodor:2009ax,deForcrand:2010ys,Aarts:2013bla}. 

One of the ways to avoid the sign problem is using the analyticity 
of the action and complexifing the field manifolds of the theory
with the complex Langevin equation \cite{Parisi:1984cs,Klauder:1983nn}.
(See also the related but distinct approach of the Lefschetz thimbles, where 
the integration contours are pushed into the complex plane \cite{Cristoforetti:2012su}.)

After promising initial results, it was noticed that the complex Langevin
equation can also deliver convergent but wrong results in some cases 
\cite{Ambjorn:1985iw,Ambjorn:1986fz}.
Also technical problems could arise which are avoided using 
adaptive step-sizes for the Langevin equation \cite{Ambjorn:1985iw,Aarts:2009dg}.
In the last decade the method has enjoyed increasing attention related 
to real-time systems \cite{Berges2005b,Berges:2006xc,Berges:2007nr,
Fukushima:2014iqa,Anzaki:2014hba}, 
as well as finite-density problems \cite{Aarts:2008rr,Aarts:2008wh,
Aarts:2010aq,Karsch:1985cb,Aarts:2011zn,Aarts:2012ft,
Pawlowski:2013pje,Pawlowski:2013gag,Pawlowski:2014ada,
Mollgaard:2013qra,Mollgaard:2014mga,Hayata:2014kra,Nishimura:2015pba}.
The method showed remarkable success in the 
case of finite density Bose gas \cite{Aarts:2008wh} or the SU(3) spin-model \cite{Karsch:1985cb,Aarts:2011zn},
but the breakdown of the method was also observed a few times \cite{Aarts:2010aq,Pawlowski:2013gag}. The theoretical understanding of the successes and the failures of the method 
has improved: it has been proved that provided a few requirements 
(some 'offline' such as the holomorphicity of the action and the 
observables, and some 'online' such as the quick decay of the field 
distributions at infinity) the method will provide correct 
results \cite{Aarts:2009uq,Aarts:2011ax}.

It has been recently demonstrated that complex Langevin simulations 
of gauge theories are made feasible using the procedure of 
gauge cooling \cite{gaugecooling,Aarts:2013uxa}, see also \cite{Nagata:2015uga}, 
which helps to reduce 
the fluctuations corresponding to the complexified gauge freedom of the 
theory. This method was first used to solve HDQCD (heavy dense QCD)
where the quarks are kept static (their spatial hopping terms 
are dropped) \cite{gaugecooling,Aarts:2013uxa}, and it has been also 
extended to full QCD using light quarks in the 
staggered \cite{Sexty:2013ica} as 
well as the Wilson formulation \cite{Aarts:2014bwa}. Gauge cooling
makes the investigation of QCD with a theta term also possible \cite{Bongiovanni:2014rna}.

In this paper we compare results 
of the reweighting approach and the Complex Langevin approach for $N_F=4$ 
and $N_F=2$ QCD using staggered fermions.

In Section \ref{clesec}. we give a brief overview of the complex Langevin method. 
In Section \ref{rewsec}. we summarize the reweighting method. 
In section \ref{ressec}. we present our numerical results comparing
the reweighting and complex Langevin simulations.
Finally, we conclude in Section \ref{concsec}.

\section{The complex Langevin equation}
\label{clesec}
The Complex Langevin equation \cite{Parisi:1984cs,Klauder:1983nn} is 
a straightforward generalization 
of the real Langevin equation \cite{Parisi:1980ys}. 
For the link variables $U_{x,\nu}$ of lattice 
QCD an update with Langevin timestep $\epsilon$  
reads \cite{PhysRevD.32.2736}:

\bea
U_{x,\nu} (\tau+\epsilon) = 
  \textrm{exp} \left[ i \sum\limits_a 
 \lambda_a ( \epsilon K_{ax\nu}  + \sqrt\epsilon \eta_{ax\nu} ) 
\right]         U_{x,\nu}(\tau),
\eea
with $\lambda_a$ the generators of the gauge group, i.e. 
the Gell-Mann matrices, and the Gaussian noise $\eta_{ax\nu}$.
 The drift force $K_{ax\nu}$ is determined from the action $S[U]$ by
\bea
K_{ax\nu}= -D_{ax\nu} S[U]
\eea
with the left derivative
\bea
D_{ax\nu}f(U) = \left. \partial_\alpha f( e^{i \alpha \lambda_a} U_{x,\nu}) \right|_{\alpha=0}.
\eea
In case the drift term is non-real the manifold of the link variables 
is complexified to SL(3,$\mathcal{C}$). The original theory is recovered 
by taking averages of the observables analytically continued to the 
complexified manifold.

For the case of QCD the action of the theory involves the 
fermionic determinant through the complex logarithm function
\bea
S_{eff}= S_{YM} - \textrm{ln det} M(\mu).
\eea
The drift term in turn is given by
\bea \label{driftterm}
 K_{ax\nu} &=& -D_{ax\nu} S_{YM} [U]  %\\ \nonumber 
 + {N_F\over 4 }\textrm Tr[ M^{-1}(\mu,U) 
  D_{ax\nu} M (\mu,U)  ],
\eea
where the second term is calculated using one CG inversion per update using 
noise vectors \cite{Sexty:2013ica}.
The action we are interested in is thus non holomorphic, and in turn this 
results in a drift term which has singularities where the fermionic measure 
$ \textrm{det} M(\mu,U)$ is vanishing. 

The theoretical understanding of the behavior of the theory with 
a meromorphic drift term is still lacking, but we 
have some observations as detailed below.
Such a drift term seems to 
lead to incorrect results in toy models if 
the trajectories encircle the origin 
frequently \cite{Mollgaard:2013qra,Greensite:2014cxa}. 
In other cases the simulations yield a correct result
in spite of a logarithm in the action \cite{Aarts:2011zn,Aarts:2012ft}.
In \cite{Nishimura:2015pba} an explicit example is presented where 
the simulations give correct results in spite of the frequent rotations 
of the phase of the measure. The condition for correctness is that the 
distribution of configurations vanishes sufficiently fast 
(faster than linearly) near the pole.

 For QCD itself we have a few indications that the poles do not affect the 
simulations at high temperatures: 
observing the spectrum of the Dirac operator \cite{Sexty:2014dxa}, comparisons 
with expansions which use a holomorphic action \cite{Aarts:2014bwa},
and the results presented in this paper. It remains to see
whether simulations in the confined phase are affected.

The 'distance' of a configuration from the original SU(3) manifold can
be monitored with the unitarity norm
\bea
\label{unitaritynorm}
{ 1 \over 4 \Omega} \sum_{x,\mu} Tr (( U_{x,\mu} U^{+}_{x,\mu} -1 )^2),
\eea
where $\Omega = N^3_s N_t $ is the volume of the lattice. In naive complex 
Langevin simulations, this distance grows exponentially, and the simulation
breaks down because of numerical problems if it gets too large.
This behavior can be countered with gauge cooling, which 
means that several gauge transformations of the enlarged manifold are 
performed in the direction of the steepest descent of the 
unitarity norm (\ref{unitaritynorm}) \cite{gaugecooling,Aarts:2013uxa}. 
With gauge cooling, the unitarity norm remains bounded at a safe 
level as long as the $\beta$ parameter of the action is not too small.
The value $\beta_{min}$ corresponds to a maximal lattice spacing, which 
seems to depend weakly on the lattice size, as can be checked 
easily for the cheaper HDQCD theory \cite{Aarts:2013nja}.

\section{Reweighting}
\label{rewsec}

In the multi-parameter reweighting approach one rewrites the partition function 
as~\cite{Fodor:2001au}:

\begin{eqnarray}
\label{eq:rew}
Z = \int {\cal D}Ue^{-S_{YM}(\beta)}\det M(\mu)= 
\int {\cal D}U e^{-S_{YM}(\beta_0)}\det M(\mu_0)
\left\{e^{-S_{YM}(\beta)+S_{YM}(\beta_0)}
\frac{\det M(\mu)}{\det M(\mu_0)}\right\},
\end{eqnarray}
where $\mu_0$ is chosen such that
the second line contains a positive definite measure 
which can be used to generate the configurations and the terms
in the curly bracket in the last line are taken into
account as an observable.
The expectation value of any observable can be then written in the form:
\begin{equation}
<O>_{\beta,\mu}=\frac{\sum O(\beta,\mu) w(\beta,\beta_0,\mu,\mu_0)}{\sum w(\beta,\beta_0,\mu,\mu_0)}
\label{rew_O}
\end{equation}
with $w(\beta,\beta_0,\mu,\mu_0)$ being the weights of the configurations defined by
the curly bracket of eqn. (\ref{eq:rew}). Note that gauge observables do
not explicitly depend on $\mu$, therefore their $\mu$ dependence comes
entirely from the weight factors. Fermionic observables, on the other hand
also explicitly depend on the chemical potential.

In this paper we use two choices for the original, positive measure ensemble.
The first choice is to use $\mu_0=0$, i.e. reweighting from zero chemical potential.
For any choice of the target $\beta, \mu$ parameters one can find
the optimal $\beta_0$ for which the fluctuation of the weights $w(\beta,\mu)$
is minimal. This corresponds to the best reweighting line as discussed in~\cite{Fodor:2002km, Csikor:2004ik,Nagata:2012pc}.
We generated configurations at $\mu=0$ for $\beta$ in the range $4.9-5.5$.
These were then used to reach the entire $\mu,\beta$ plane via multi-parameter 
reweighting. Our second choice is to use the phase quenched ensemble, i.e. replacing 
$\det M(\mu_0)$ by $|\det M (\mu)|$ in eqs.~(\ref{eq:rew}) and (\ref{eq:root}). In this case the reweighting factor 
contains only the phase of the determinant.

For staggered fermions an additional rooting is required, for $N_F$ flavors the
weights become
\begin{equation}
\label{eq:root}
w(\beta,\beta_0,\mu,\mu_0)= e^{-S_{YM}(\beta)+S_{YM}(\beta_0)}
\left[ \frac{\det M(\mu)}{\det M(\mu_0)} \right] ^{N_F/4}
\end{equation}
Since for $N_F<4$ a fractional power is taken which has cuts on the complex plane
it is important to choose these cuts such that the weights are analytic for
real $\mu$ values. This can be achieved by expressing $\det M(\mu)$ analytically
as a function of $\mu$ as discussed in~\cite{Gibbs:1986hi,Fodor:2001pe}.

\section{Results}
\label{ressec}

We use the Wilson plaquette action for the gauge sector of the theory and 
unimproved staggered fermions with $N_F=4$ flavors if not otherwise noted.
We have used three different lattice sizes for 
this study, $ 8^3 \times 4$, $ 12^3 \times 6$ and $16^3 \times 8 $, all
 having the aspect ratio $ L_s / L_t=2$.

Our main observables are the plaquette averages, the spatial average 
of the trace of the Polyakov loop 
\bea \label{polyakovloop}
 \sum_x \textrm{Tr} P(x) /N^3_s, \qquad       P(x) =  \prod_{i=1..N_t} U_4(x,i)
\eea
and its inverse $ \sum_x \textrm{Tr} P^{-1}(x) /N^3_s $ , the chiral condensate $ \langle \bar \psi \psi \rangle $ 
and the fermionic 
density $n$ defined as 
\bea
\langle \bar\psi \psi \rangle = {1\over \Omega} \left\langle {  \partial \ln Z \over \partial m } \right\rangle , \qquad
 n={1\over \Omega} \left\langle
{\partial \ln Z \over \partial \mu }\right\rangle,
\eea
with $ \Omega$ the volume of the space-time lattice.
We are also interested in the average phase of the fermion determinant, which 
measures the severity of the sign problem
\bea \label{phaseavr}
\langle e^{2 i \varphi} \rangle = \left\langle {\det M( \mu) \over \det M(-\mu)} \right\rangle .
\eea

\begin{table}
\begin{center}
\begin{tabular} 
{|c|c|c|c|}\hline
$ \beta $ & $a m_q$  & $ a m_\pi $ & $a$ (fm) \\
\hline

4.80&0.01&$ 0.2458 \pm 0.0007 $ & $ 0.3355 \pm 0.0001  $ \\ 
4.85&0.01&$ 0.2480 \pm 0.0007 $ & $ 0.3315 \pm 0.0001 $ \\ 
4.90&0.01&$ 0.2506 \pm 0.0009 $ & $ 0.3258 \pm 0.0003  $\\ 
4.95&0.01&$ 0.2533 \pm 0.0009 $ & $ 0.3174 \pm 0.0003 $\\ 
% 5.00&0.01&$ 0.2577 \pm 0.0010 $ & $ 0.3045 \pm 0.0003  $\\ 

5.00 & 0.01 & $0.2596 \pm 0.0005$ & $ 0.2892 \pm  0.0001 $ \\
5.05 &0.01&$ 0.2679 \pm 0.0008 $ & $ 0.2773 \pm 0.0006  $\\ 

5.10& 0.01 & $ 0.2870 \pm 0.0010$ & $ 0.1890 \pm 0.0006 $ \\
%5.10&0.01&$ 0.2888 \pm 0.0010 $ & $ 0.202 \pm 0.001 $\\ 
5.15&0.01&$ 0.3014 \pm 0.0012 $ & $ 0.1410 \pm 0.001 $\\ 
5.20 & 0.01 & $ 0.2957 \pm   0.0018$ & $ 0.1123 \pm 0.0006 $ \\
%5.20&0.01&$ 0.3086 \pm 0.0021 $ & $ 0.111 \pm 0.001  $\\ 
5.25&0.01& $ 0.2918 \pm 0.0021$ & $ 0.0957 \pm 0.0006$ \\
5.40&0.01&$ 0.2456 \pm 0.0012 $ &  $0.0652\pm 0.0006 $ \\

\hline 

\hline
4.80&0.02&$ 0.3455 \pm 0.0006 $ & $ 0.3362 \pm   0.0001 $\\
4.85&0.02&$ 0.3491 \pm 0.0005 $ & $0.3323 \pm 0.0001 $\\ 
4.90&0.02&$ 0.3520 \pm 0.0007 $ & $0.3270 \pm 0.0001 $\\ 
4.95&0.02&$ 0.3568 \pm 0.0006 $ & $0.3197 \pm 0.0001 $\\ 
5.00&0.02&$ 0.3629 \pm 0.0007 $ & $0.3084 \pm 0.0002 $\\ 
5.05&0.02&$ 0.3726 \pm 0.0005 $ & $0.2878 \pm 0.0004 $\\ 
5.10&0.02&$ 0.3898 \pm 0.0008 $ & $0.2420 \pm 0.0009 $\\ 
5.15&0.02&$ 0.4073 \pm 0.0006 $ & $0.1751 \pm 0.0008 $\\ 
5.20&0.02&$ 0.4125 \pm 0.0011 $ & $0.1341 \pm 0.0012 $\\ 
\hline

5.00 & 0.03 & $0.4392 \pm   0.0002 $&$ 0.2986 \pm 0.0001 $\\
5.10 & 0.03 & $0.4635  \pm 0.0004$&$   0.2419 \pm 0.0002$\\
5.20 & 0.03 & $ 0.4905 \pm 0.0008 $&$  0.1467 \pm 0.0004$\\
5.25 &0.03 & $0.4862\pm  0.0007 $&$ 0.1237 \pm 0.0003 $\\
5.40 & 0.03 &$ 0.4407 \pm 0.0009$&$  0.0849 \pm 0.0004 $\\

\hline

4.80&0.05&$ 0.5413 \pm 0.0003 $ & $ 0.3379 \pm 0.0001   $ \\ 
4.85&0.05&$ 0.5442 \pm 0.0005 $ & $ 0.3345 \pm 0.0001  $ \\ 
4.90&0.05&$ 0.5480 \pm 0.0004 $ & $ 0.3301 \pm 0.0001  $ \\ 
4.95&0.05&$ 0.5530 \pm 0.0005 $ & $ 0.3241 \pm 0.0001  $ \\ 
%5.00&0.05&$ 0.5583 \pm 0.0004 $ & $ 0.3163 \pm 0.0002 $ \\ 
 5.00& 0.05& $ 0.5588 \pm 0.0002 $&$ 0.3045 \pm 0.0001 $ \\

5.05&0.05&$ 0.5678 \pm 0.0004 $ & $ 0.3042 \pm 0.0002 $ \\ 
%5.10&0.05&$ 0.5801 \pm 0.0005 $ & $ 0.2822 \pm 0.0003 $ \\ 
5.10 &0.05 & $ 0.5784 \pm 0.0003  $&$ 0.2664 \pm  0.0002 $\\

5.15&0.05&$ 0.5961 \pm 0.0006 $ & $ 0.2426 \pm 0.0008 $ \\ 
%5.20&0.05&$ 0.6094 \pm 0.0006 $ & $ 0.1894 \pm 0.0009 $ \\ 
5.20 &0.05 & $ 0.6100 \pm 0.0005 $ & $ 0.1783 \pm  0.0003 $\\
5.25& 0.05 &$ 0.6144 \pm 0.0006 $ & $  0.1465 \pm  0.0004 $\\

\hline
\end{tabular}
\end{center}
\caption{Pion masses and lattice spacings for different $\beta$ values 
and bare quark masses,
measured on $12^3 \times 24 $, $16^3\times 32$  and $24^3 \times 48$ lattices with $N_f=4$.
Statistical errors are indicated.
   }
\label{tab:masses}
\end{table}

We perform the complex Langevin simulations using adaptive step-size, with 
a control parameter which puts the typical step-sizes in the 
range $ \epsilon \approx 10^{-5} - 5 \times 
10^{-5} $. 
Using such small step sizes allows us to avoid having to take 
the $\epsilon \rightarrow 0$ limit as the results are in the zero 
Langevin step limit within errors. 
We use initial conditions on the $SU(3)$ manifold, and 
allow $ \tau =10 -30$ Langevin time for thermalization, after which we perform 
the measurements for an other  $ \tau =10 -30$ Langevin time. 
We checked that proper thermalization is reached by observing that halving 
the thermalization time leads to consistent results.

We have determined the pion masses as well as the lattice spacing 
using the $w_0$ scale as proposed in \cite{Borsanyi:2012zs}
for several quark masses, see in Table~\ref{tab:masses}.
One sees that choosing the quark masses $ma=0.05 $ for the $N_t=4$ lattice, 
$ma=0.02$ for the $N_t=6$ lattice 
and $ma=0.01$ for the $N_t=8$ lattice,
in the vicinity of the critical 
temperature we have $ m_{\pi} / T_c \approx 2.2 - 2.4 $.
We have additionally investigated the $N_t=8$ lattice with $ am=0.05$, which corresponds to the rather heavy pion mass of $ m_{\pi} / T_c \approx 4.8 $.

\subsection{Reweighting from $\mu=0$}

First we have tested the theory at a fixed $\beta=5.4$ at $N_t=4$ as a function 
of $\mu$, which is well above the deconfinement transition 
which at $\mu=0$ and $m=0.05$ is at $ \beta_c \approx 5.04 $.

\begin{figure}
\begin{center}
 \epsfig{file=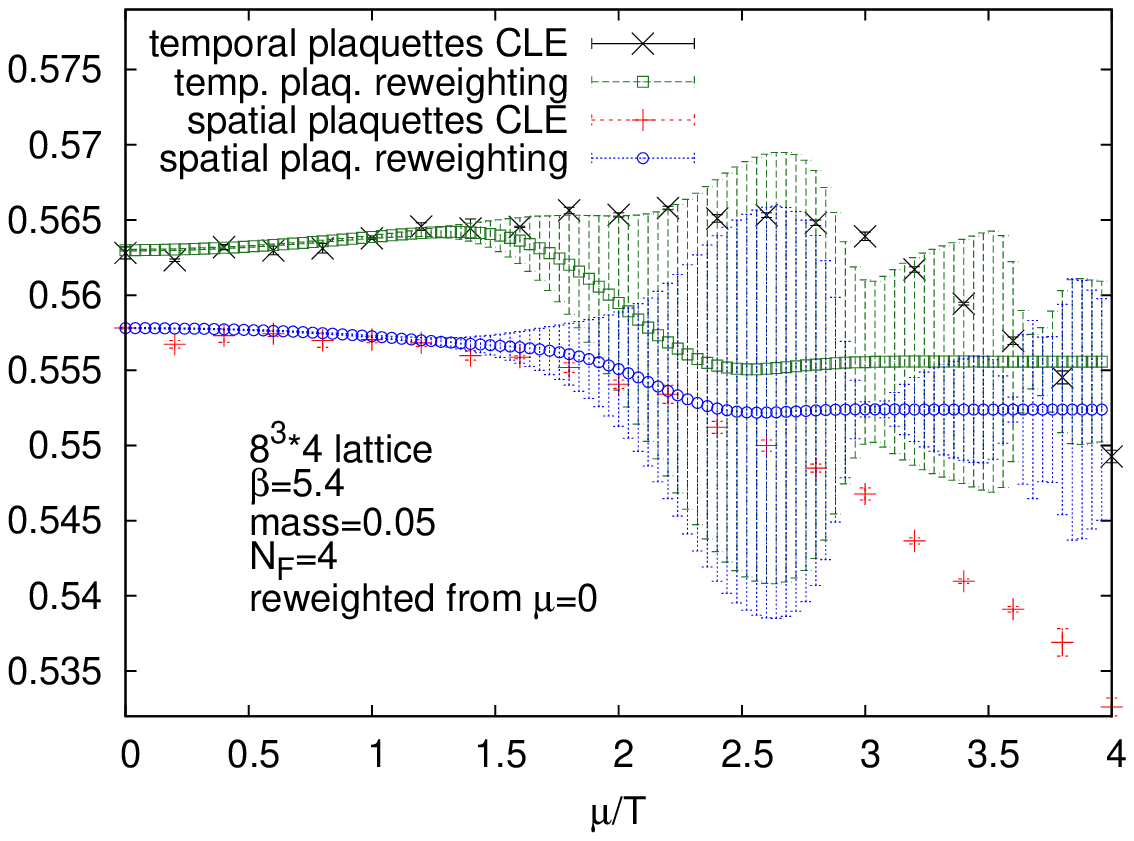, width=8cm}
 \epsfig{file=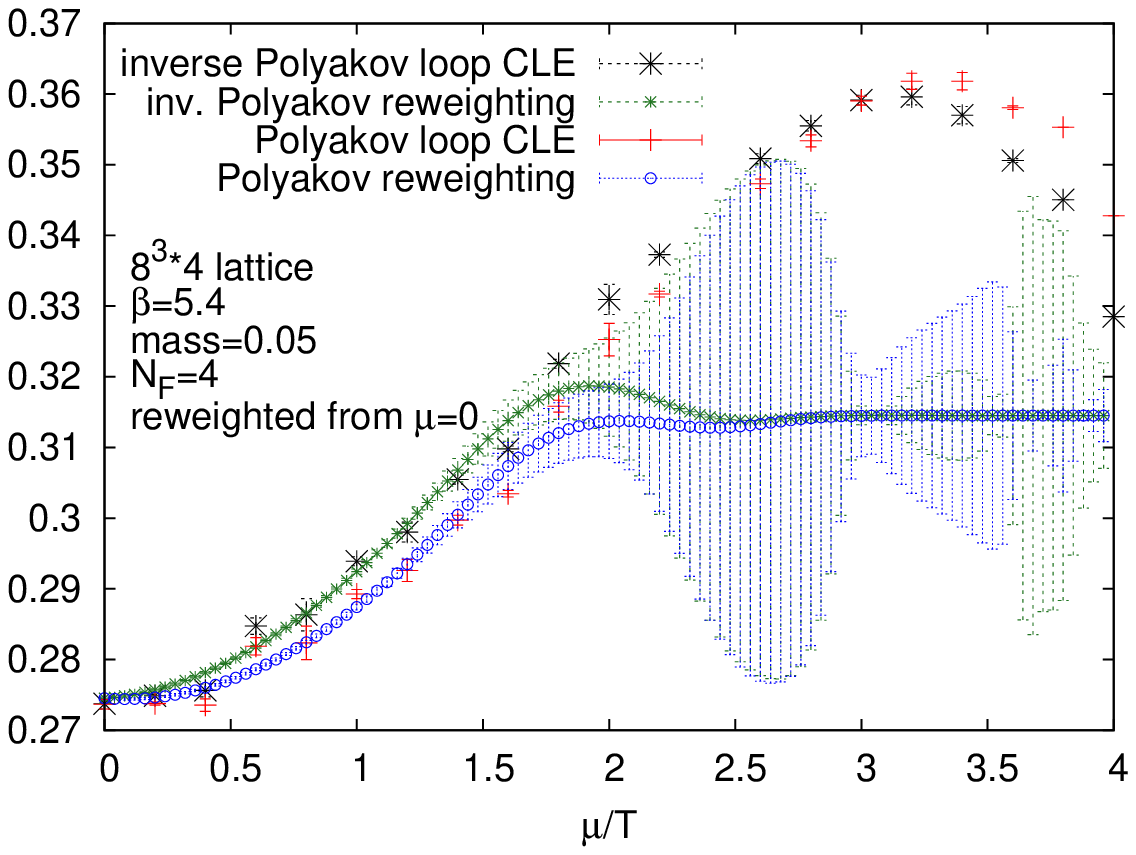, width=8cm}
 \caption{Comparison of plaquette averages and Polyakov loops and inverse Polyakov loops (defined in and below eq.(\ref{polyakovloop}))   calculated with CLE and reweighting from the $\mu=0$ ensemble. }
\label{fig-nt4-plaq} 
\end{center}
\end{figure}

\begin{figure}
\begin{center}
 \epsfig{file=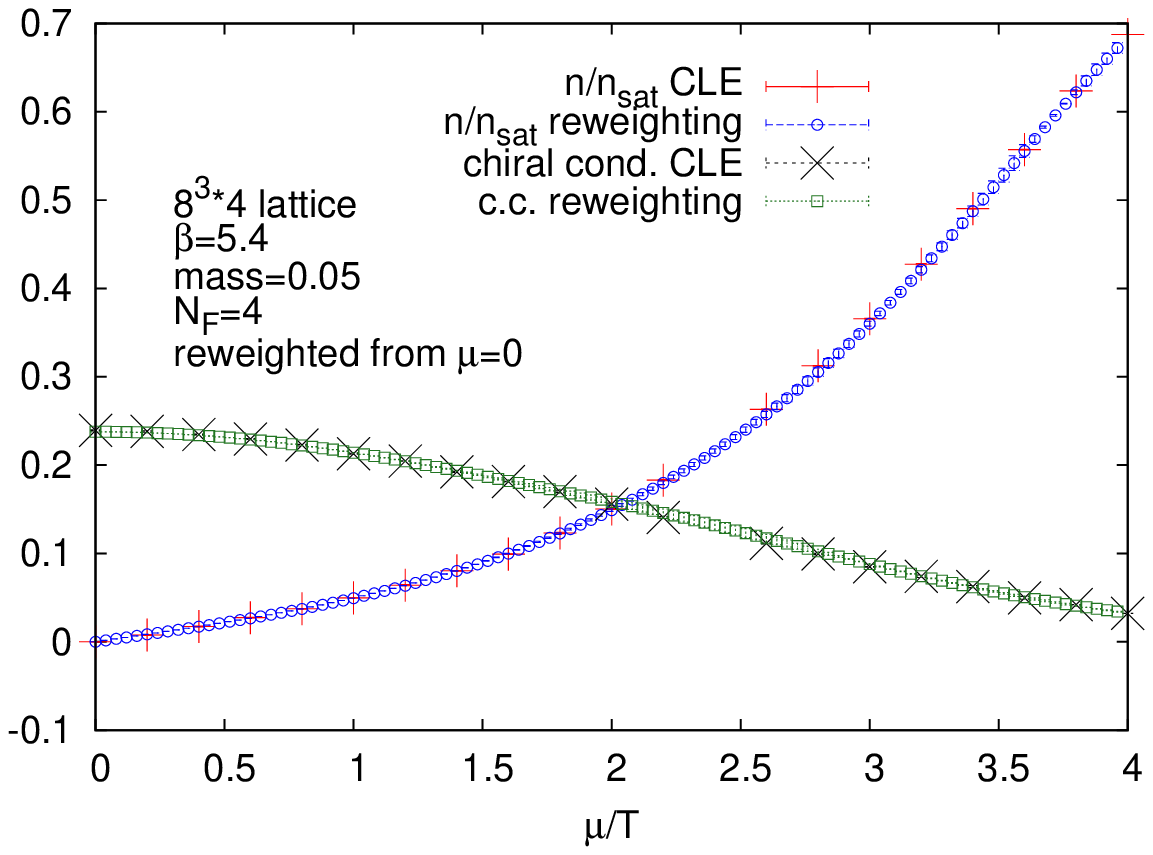, width=8cm}
 \epsfig{file=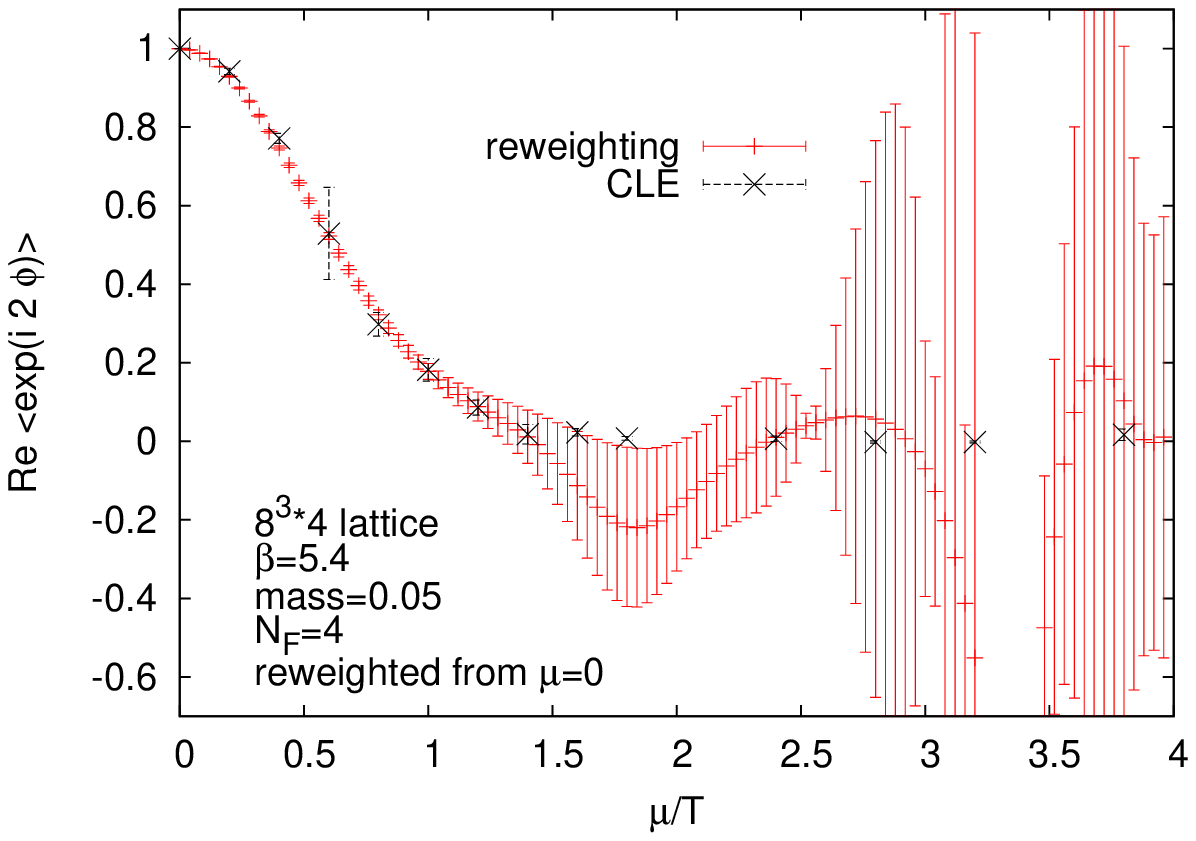, width=8cm}
 \caption{Comparison of the chiral condensate and fermionic density 
as well as the phase average (\ref{phaseavr})
calculated with CLE and reweighting from the $\mu=0$ ensemble. }
\label{fig-nt4-fermion} 
\end{center}
\end{figure}

In Fig.~\ref{fig-nt4-plaq} we show the 
comparison of the gauge observables: plaquette averages and Polyakov loops.
We generated $O(10^4)$ independent configurations in the $\mu=0$ ensemble
with the usual HMC algorithm (using every 50th configuration of the 
Markov chain), and we calculated the reweighting as 
detailed in section \ref{rewsec}. One notes that the reweighting 
performs well for small chemical potentials $\mu/T < 1-1.5 $, where 
there is a nice agreement between reweighting and CLE.
The errors of the reweighting approach start to grow large as one 
increases $\mu$ above $1.5 T$, where the average of the 
reweighting is dominated by a few configurations. This is the manifestation 
of the overlap problem: the ensemble we have sampled has typical configurations
which are not the typical configurations of the ensemble we wish to study.

Next we turn to the fermionic observables: chiral condensate, fermionic density in 
Fig.~\ref{fig-nt4-fermion}. One notes that the reweighting of these quantities is possible
to much higher values of $\mu/T$. This is the consequence of their explicit dependence 
on $\mu$, which dominates their change as the chemical potential is changed. 
This is in contrast to the gauge observables in Fig.~\ref{fig-nt4-plaq}, where the change
is given entirely by the change in the measure of the path integral.
The downward turn of the Polyakov loop and its inverse around 
$ \mu/T = 3 $ is the result of the phenomenon of saturation: at this 
chemical potential half of all of the available fermionic 
states on the lattice are filled, as visible on Fig.~\ref{fig-nt4-fermion}. 
This lattice artifact can also be observed 
with static quarks \cite{gaugecooling}, and even in the strong 
coupling expansion \cite{DePietri:2007ak}.

\begin{figure}
\begin{center}
 \epsfig{file=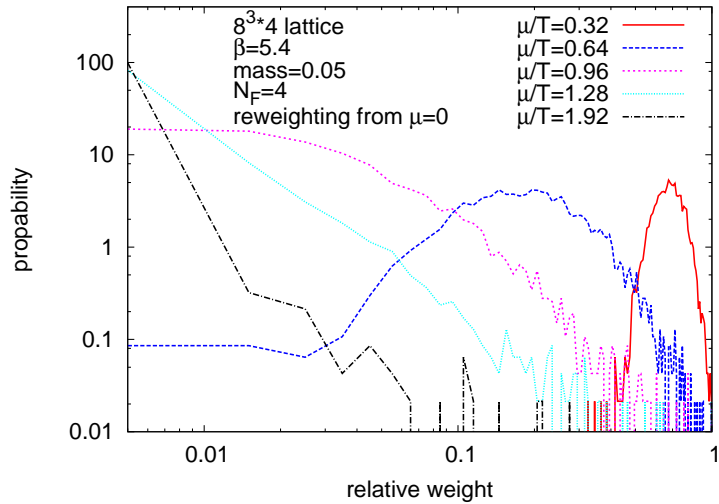, width=10cm}
 \caption{The histogram of the relative weight of the configurations 
for different chemical potentials. }
\label{fig-overlap} 
\end{center}
\end{figure}

Finally, the average phase factor in Fig.~\ref{fig-nt4-fermion} is 
a good indicator of the 
severeness of the sign problem in the theory. One sees that the average phase in the region $ \mu/T > 1.5-2$ indeed gets very small. 
Note that to see agreement between CLE and reweighting one has to be careful
to choose the observable to be the analytic continuation of an observable 
on the SU(3) manifold. In this case we define the phase factor from
the analytic continuation of the determinants, as written in (\ref{phaseavr}).

 In Fig.~\ref{fig-overlap} we show the histogram of the absolute value 
of the weights of the configurations normalized by the biggest weight
in the ensemble. This illustrates the overlap problem: the 'further' 
one tries to reweight from the original ensemble, the less and less will 
be the contribution of an average configuration to the average, which 
becomes dominated by very few configurations.  Thus the fluctuations of 
the result become larger, and even the errorbars are not 
reliable as the distribution of the observables becomes non-Gaussian. 
As we show below,
this situation improves if one chooses an ensemble 'closer' to the target 
ensemble: in this case taking the phasequenched ensemble ($ | det M(\mu) |$)
instead of the zero $\mu$ ensemble.

%\begin{figure}
%\begin{center}
% \epsfig{file=plaqs_chk_nt6_m02_b53.eps, width=8cm}
% \epsfig{file=pol_chk_nt6_m02_b5.3.eps, width=8cm}
% \caption{Comparison of plaquette averages and Polyakov loops calculated with CLE and reweighting from the $\mu=0$ ensemble on a $12^3*6$ lattice. }
%\label{fig-nt6-plaq} 
%\end{center}
%\end{figure}

%\begin{figure}
%\begin{center}
% \epsfig{file=density_nt6_m02_b53.eps, width=8cm}
% \caption{Comparison of the chiral condensate and fermionic density 
%as well as the phase average (\ref{phaseavr})
%calculated with CLE and reweighting from the $\mu=0$ ensemble. }
%\label{fig-nt6-fermion} 
%\end{center}
%\end{figure}
%The reweighting from $\mu=0$ behaves similarly at $N_t=6$, as 
%depicted in Fig.~\ref{fig-nt6-plaq} for the gluonic 
%observables. As discussed earlier, we have changed the quark mass 
%to $ma=0.02$ to keep the pion mass approximately constant in physical 
%units. We have choosen $\beta=5.3$ which is well above the deconfinement
%transition for $N_t=6$ and $N_F=4$.
%One observes a weaker dependence on the chemical potential, and reweighting
%seems to work again only up to $ \mu/T < 1- 1.5$.
%In Fig.~\ref{fig-nt6-fermion} we plot the fermionic density. One observes that 
%in contrast to the $N_t=4$ case, there is a slight discrepancy 
%for $\mu/T > 2 $. The appearence of such a discrepancy can be already
%expected from the typical growth of the errors for e.g. the 
%spatial plaquettes (c.f. Fig.~\ref{fig-nt6-plaq} left panel).

\begin{figure}
\begin{center}
 \epsfig{file=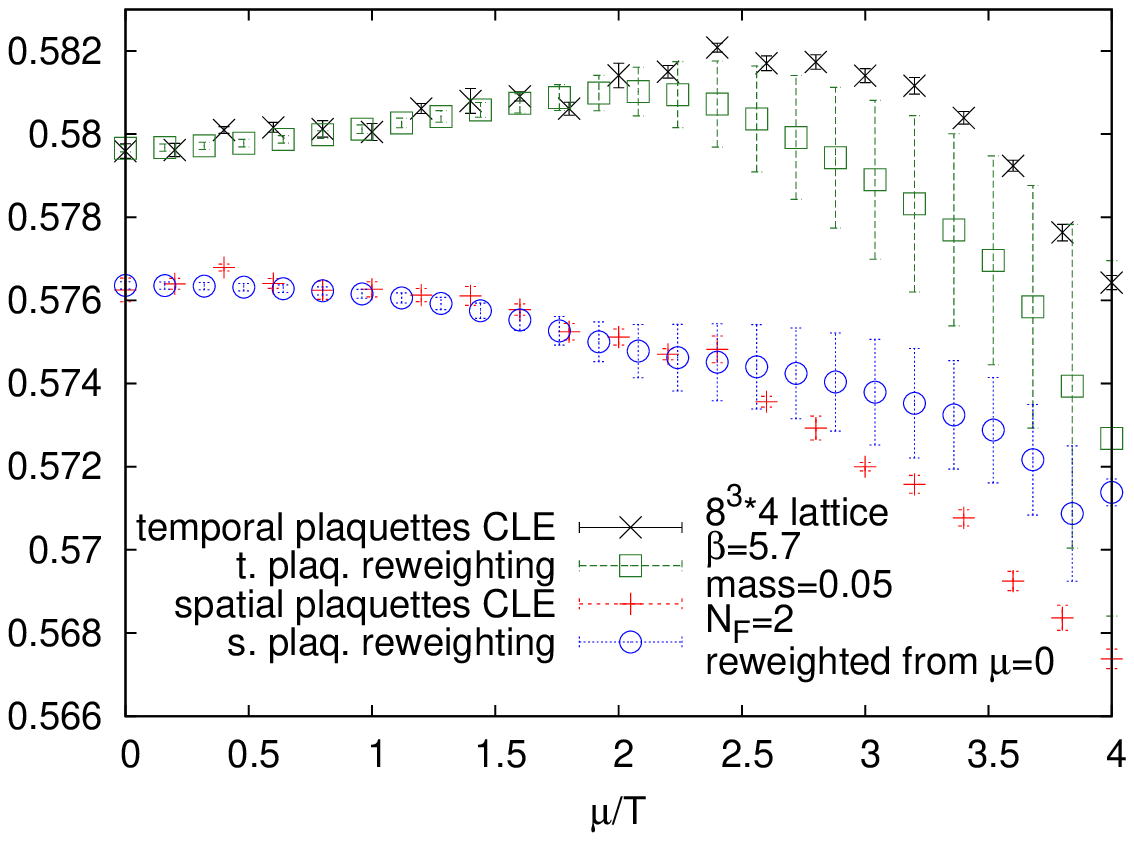, width=8cm}
 \epsfig{file=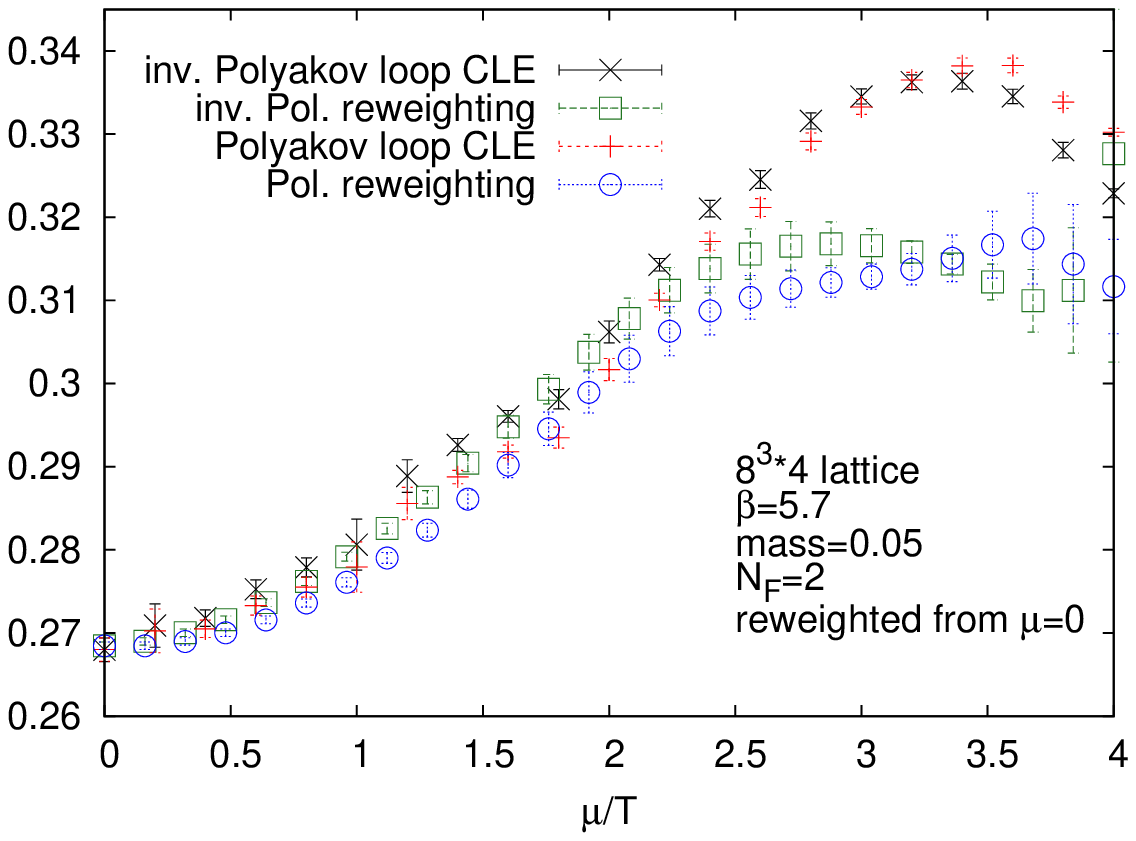, width=8cm}
 \caption{Comparison of plaquette averages and Polyakov loops calculated with CLE and reweighting from the $\mu=0$ ensemble. We use $N_F=2$ fermion flavours for the data presented in this plot. }
\label{fig-nt4-plaq-nf2} 
\end{center}
\end{figure}

In Fig.~\ref{fig-nt4-plaq-nf2} we use a theory with $N_F=2$ flavors of 
fermions, by taking the square root of the 
staggered fermion determinant. 
We perform reweighting from the $\mu=0$ ensemble using 
$\approx 1700$ configurations.
To maintain analyticity, in the 
reweighting procedure one must make sure that no cut of the complex 
square root function is crossed while the chemical potential is changed.
In the complex Langevin simulations the rooting is implemented simply by 
multiplying the fermion drift terms with an 
appropriate factor \cite{Sexty:2013ica}. We observe good agreement 
for small values $\mu/T$, similarly to case of the $N_F=4$ theory, indicating 
that the effect of rooting is the same in these different approaches.

\subsection{Reweighting from the phasequenched ensemble}

\begin{figure}
\begin{center}
 \epsfig{file=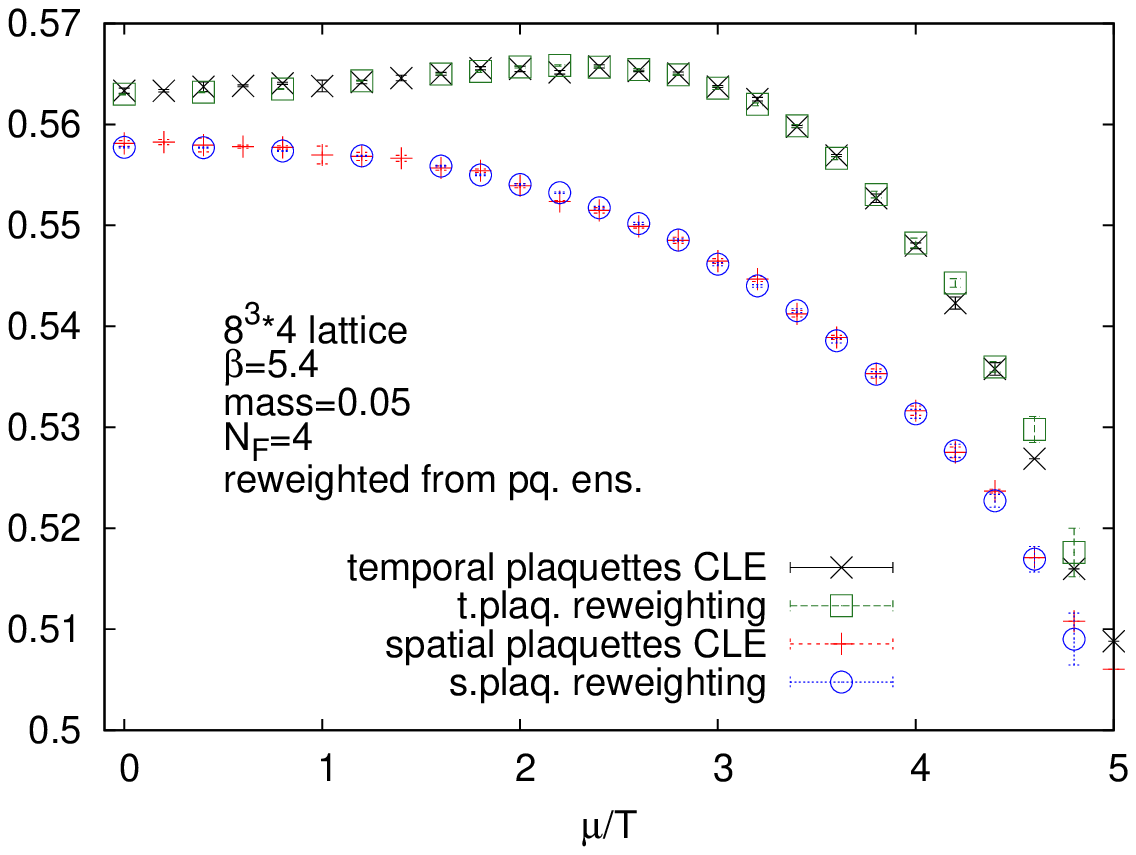, width=8cm}
 \epsfig{file=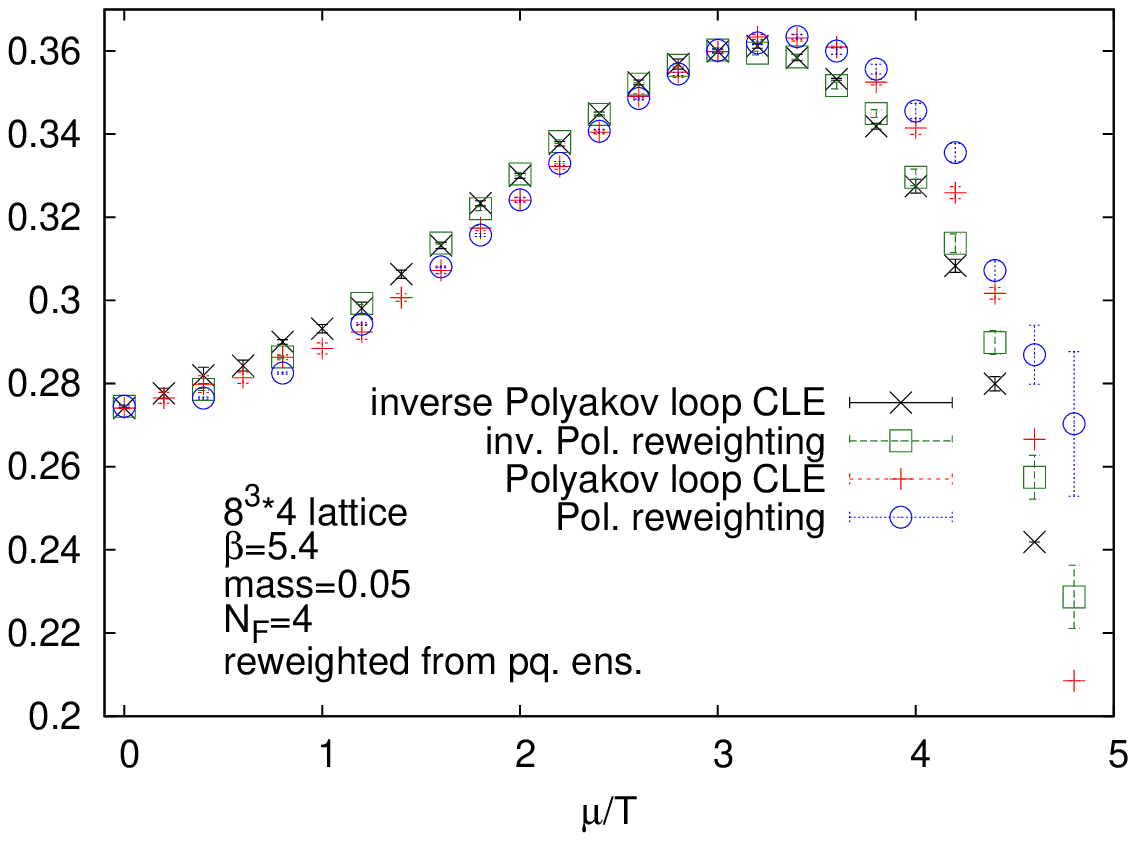, width=8cm}
 \caption{Comparison of the plaquette averages and Polyakov loops 
as a function of $\mu$ at a fixed $\beta=5.4$ 
calculated with CLE and reweighting from the phasequenched ensemble. }
\label{fig-nt4-pqrew} 
\end{center}
\end{figure}

We have investigated the efficiency of reweighting from the 'phasequenched' 
ensemble. On Fig.~\ref{fig-nt4-pqrew} we show the comparison of the plaquette 
averages as well as the Polyakov loop averages. 
We have used about 4000-5000 independent configurations at $N_t=4$ for each $\mu$ value. One notes that the agreement
is much better when compared to the reweighting from the $\mu=0$ ensemble, 
also for higher $ \mu/T$ values (compare with Fig.~\ref{fig-nt4-plaq}).
Note that this comparison is in the deconfined phase, therefore 
no phase transition corresponding to the pion condensation is expected 
in the phasequenched ensemble, makeing reweighting easier.
For the $\beta=5.4$ value used for these 
plots, the complex Langevin simulation breaks down in the saturation 
region $ \mu/T >5 $ (not shown in the plots), also signaled by a large 'skirt'
of the distributions (meaning a slow, typically power law decay) 
and the disagreement of the reweighting and 
CLE simulations, most detectable in the plaquette averages.

\subsection{Comparisons as a function of $\beta$}

\begin{figure}
\begin{center}
 \epsfig{file=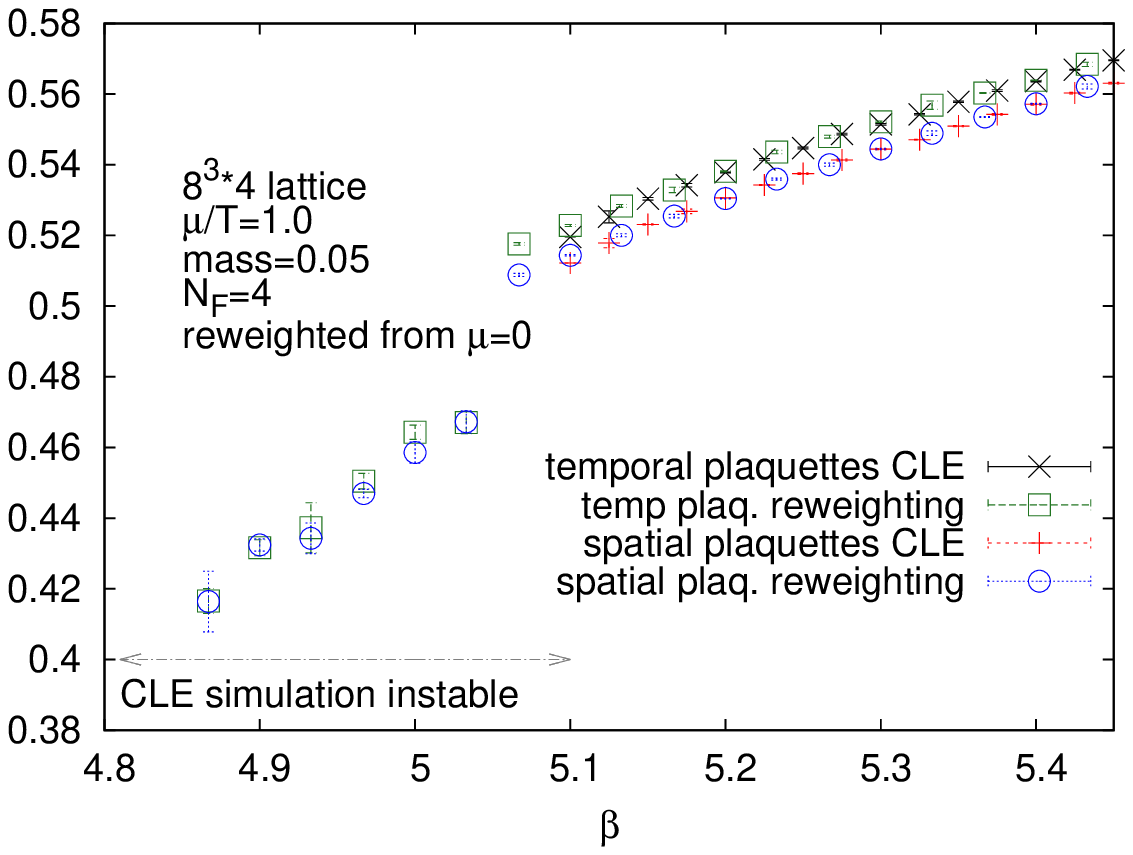, width=8cm}
 \epsfig{file=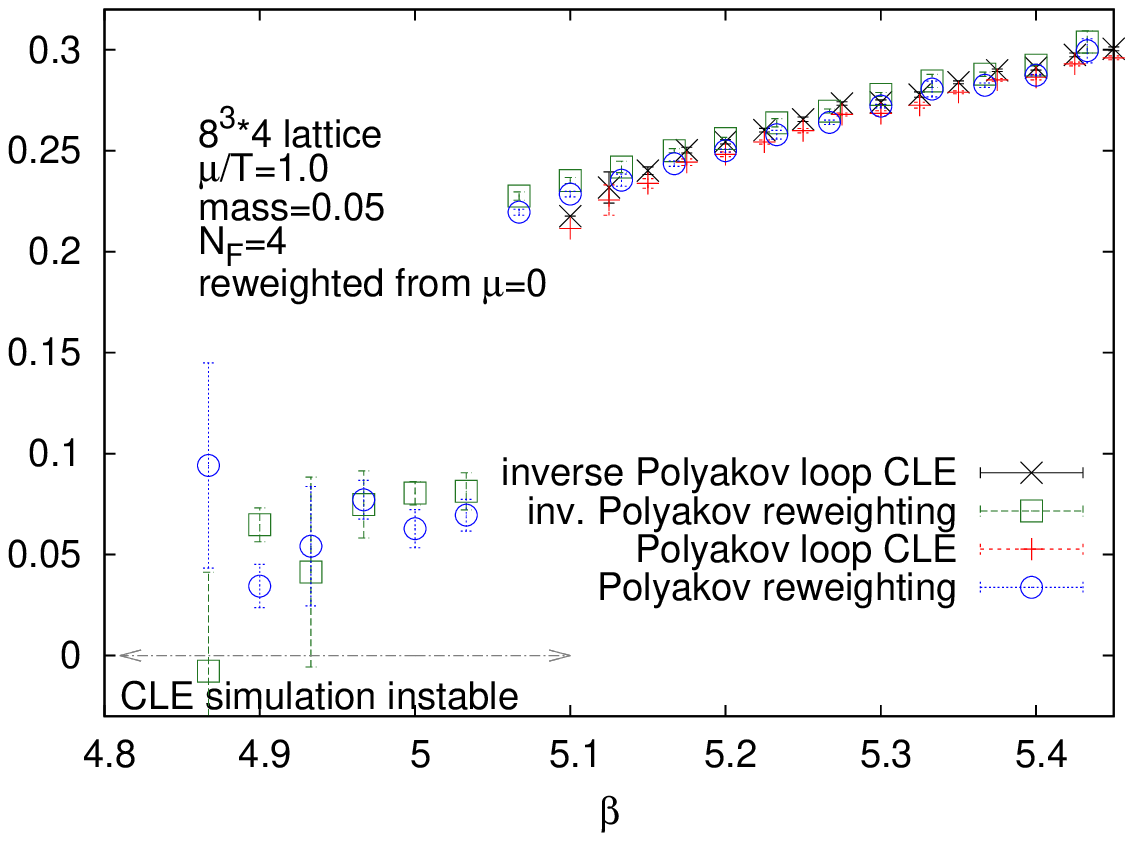, width=8cm}
 \caption{Comparison of the plaquette averages and Polyakov loops 
as a function of the 
$\beta$ parameter at a fixed $\mu/T=1$ 
calculated with CLE and reweighting from the $\mu=0$ ensemble. }
\label{fig-nt4-vertical} 
\end{center}
\end{figure}

\begin{figure}
\begin{center}
 \epsfig{file=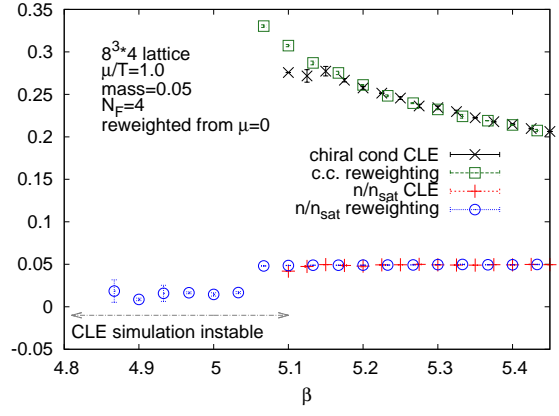, width=8cm}
 \caption{Comparison of the chiral condensate and the fermionic density 
as a function of the 
$\beta$ parameter at a fixed $\mu/T=1$ 
calculated with CLE and reweighting from the $\mu=0$ ensemble. }
\label{fig-nt4-vertferm} 
\end{center}
\end{figure}

\begin{figure}
\begin{center}
 \epsfig{file=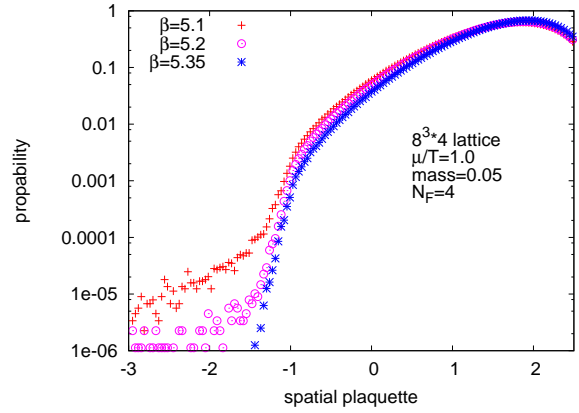, width=8cm}
 \caption{The histograms of the spatial plaquette variable 
in the CLE simulation, measured at various 
$\beta$ values corresponding to Figs~\ref{fig-nt4-vertical} and 
\ref{fig-nt4-vertferm}. 
 }
\label{fig-szoknya} 
\end{center}
\end{figure}

Next we have investigated the appearance of a discrepancy of the 
CLE and reweighting results at smaller $\beta$ values arising from 
a 'skirt' of the complexified distributions \cite{gaugecooling,Aarts:2013uxa}. 
In Fig.~\ref{fig-nt4-vertical} we compare reweighting and CLE as a function
of the $\beta$ parameter at fixed $\mu/T=1$ on an $8^3*4$ lattice. One observes that the reweighting is nicely reproduced by the Complex Langevin simulations 
as long as $\beta > 5.10- 5.15$. Below this limit the distributions 
develop a long skirt and CL simulations become instable, 
also signaled by large unitarity norm and the 
conjugate gradient algorithm 
(needed for the calculation of the drift terms in the CLE) 
failing to converge.
Similar behavior is detected on the fermionic observables in 
Fig.~\ref{fig-nt4-vertferm}.
This behavior has been observed also in 
HDQCD simulations \cite{gaugecooling,Aarts:2013nja}, where a limit value 
$\beta_\textrm{min} = 5.6-5.7 $ was seen independent 
of the value of $N_t \ge 6 $, and $\beta_{min}$ was slightly 
smaller for $N_t=4 $.  This minimal $\beta$ parameter corresponds to 
a maximal lattice spacing $ a_{max} \approx 0.2\textrm{fm}$ in HDQCD.
Apparently the limiting $\beta$ value is different in full QCD, but it 
turns out that the corresponding lattice spacing is roughly equal for 
$N_F=4$ with $am=0.05$: $ a_{max} \approx 0.2-0.25\ \textrm{fm}$.
 This breakdown is also visible on histograms of various observables.
In Fig.~\ref{fig-szoknya} we show the histograms of the spatial plaquettes
at various $\beta$ values. One notices that the 'skirt' of the distribution
is indeed large at $\beta=5.1$, where the CLE breaks down. Although a 
small skirt is also 
present at $\beta=5.2$, it is not visited frequently enough to change the 
averages noticeably. 

\begin{figure}
\begin{center}
 \epsfig{file=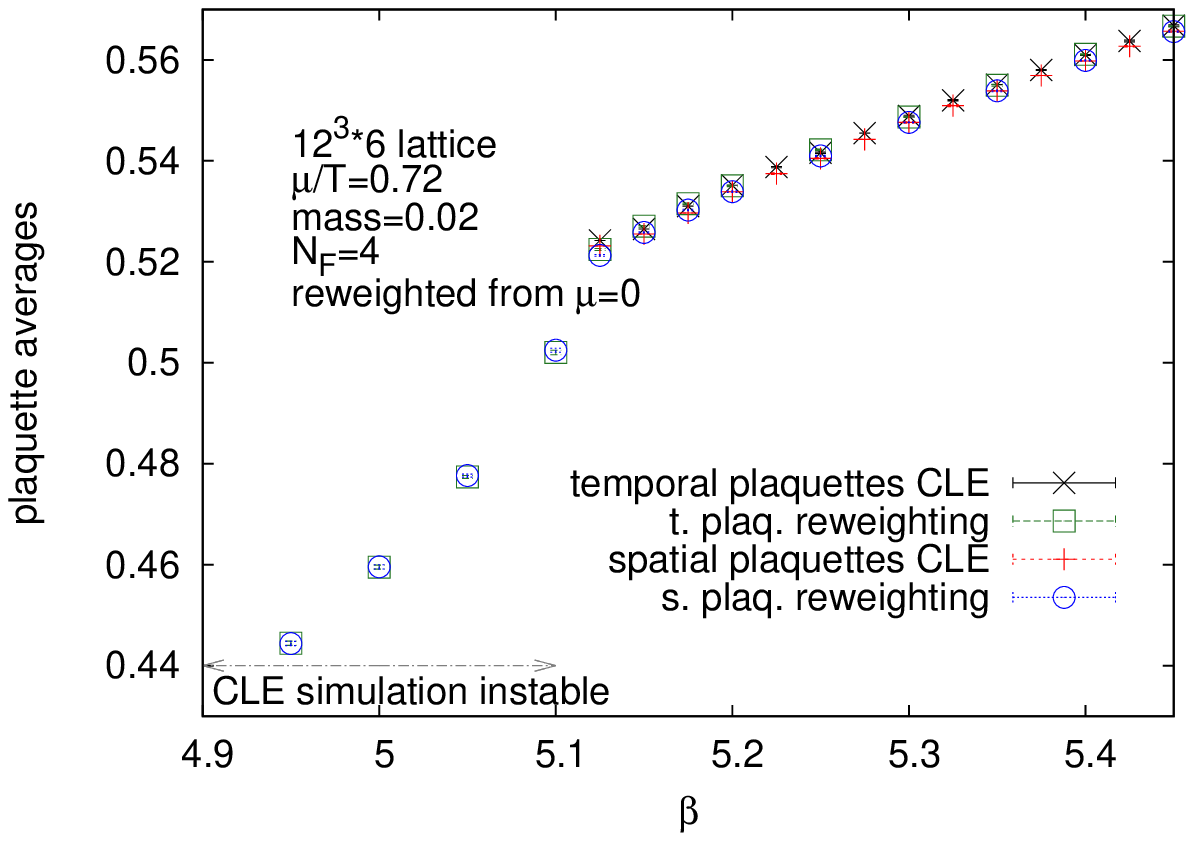, width=8cm}
 \epsfig{file=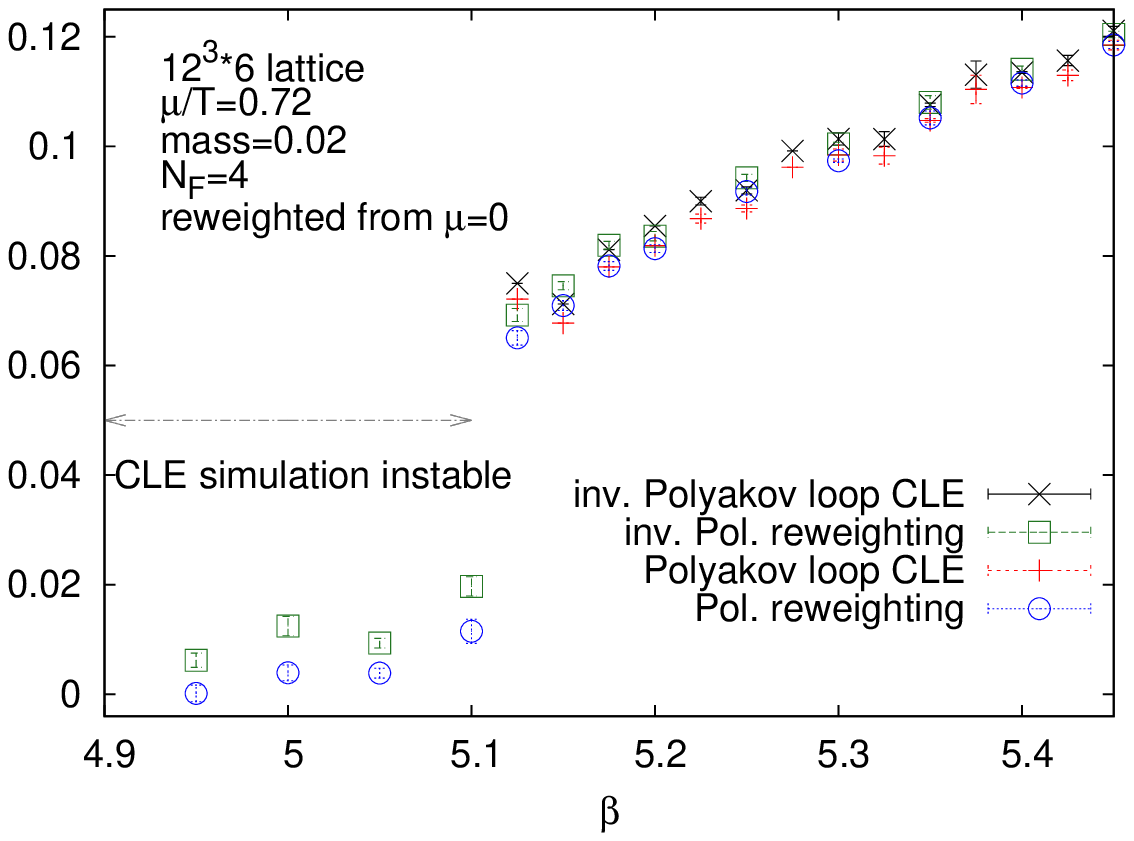, width=8cm}
 \caption{Comparison of the plaquette averages and Polyakov loops 
as a function of the 
$\beta$ parameter at a fixed $\mu/T=0.72$ 
calculated with CLE and reweighting from the $\mu=0$ ensemble on
$12^3*6$ lattices.
 }
\label{fig-nt6-vertical} 
\end{center}
\end{figure}

A similar behavior is observed on the finer $ 12^3 \times 6 $ lattice,
as depicted in Fig.~\ref{fig-nt6-vertical}.  We used 200-300 configurations 
for the reweighting procedure on $N_t=6$ lattices at every 
$\beta$ value.
We observe a limiting $\beta_{min} \approx 5.15$ 
corresponding to $ a_{max} \approx 0.15 \textrm{ fm} $
which at $N_t=6$ allows simulations right 
down to the transition temperature, but not below.

\begin{figure}
\begin{center}
 \epsfig{file=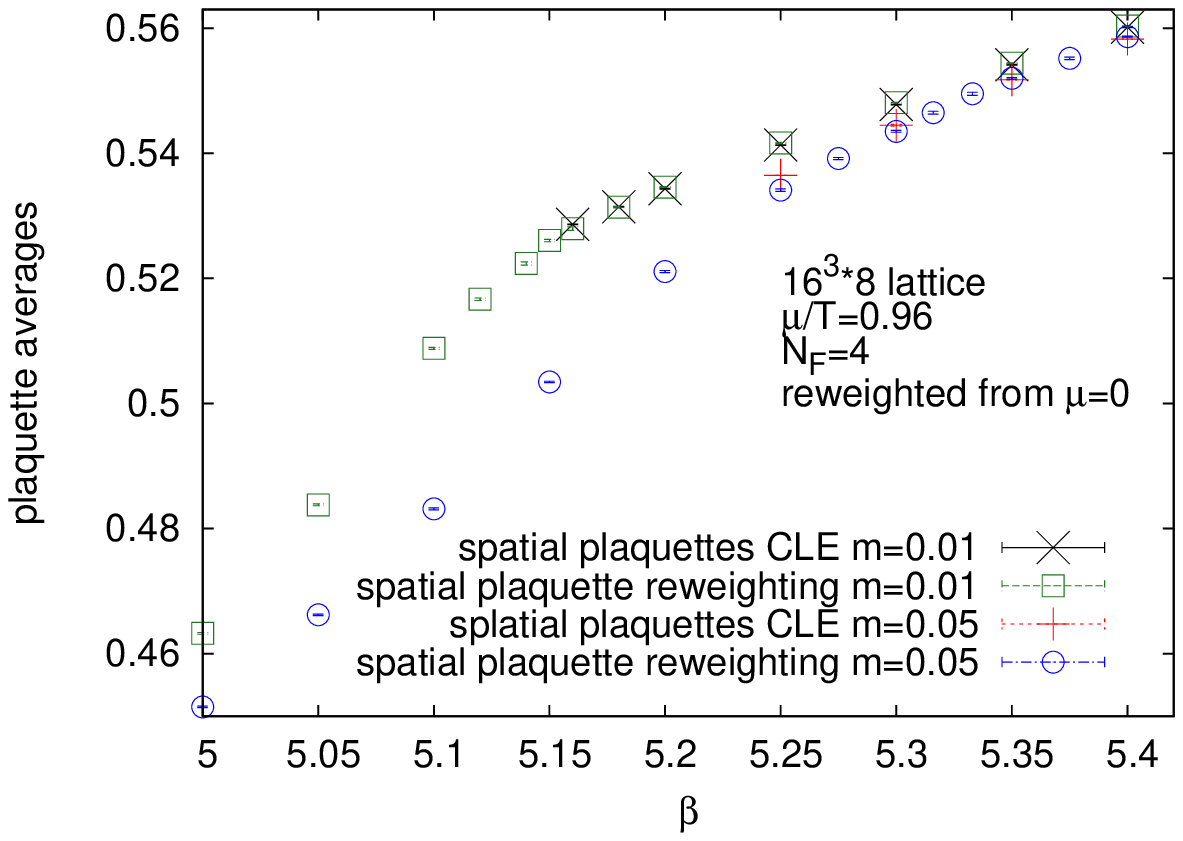, width=8cm}
 \epsfig{file=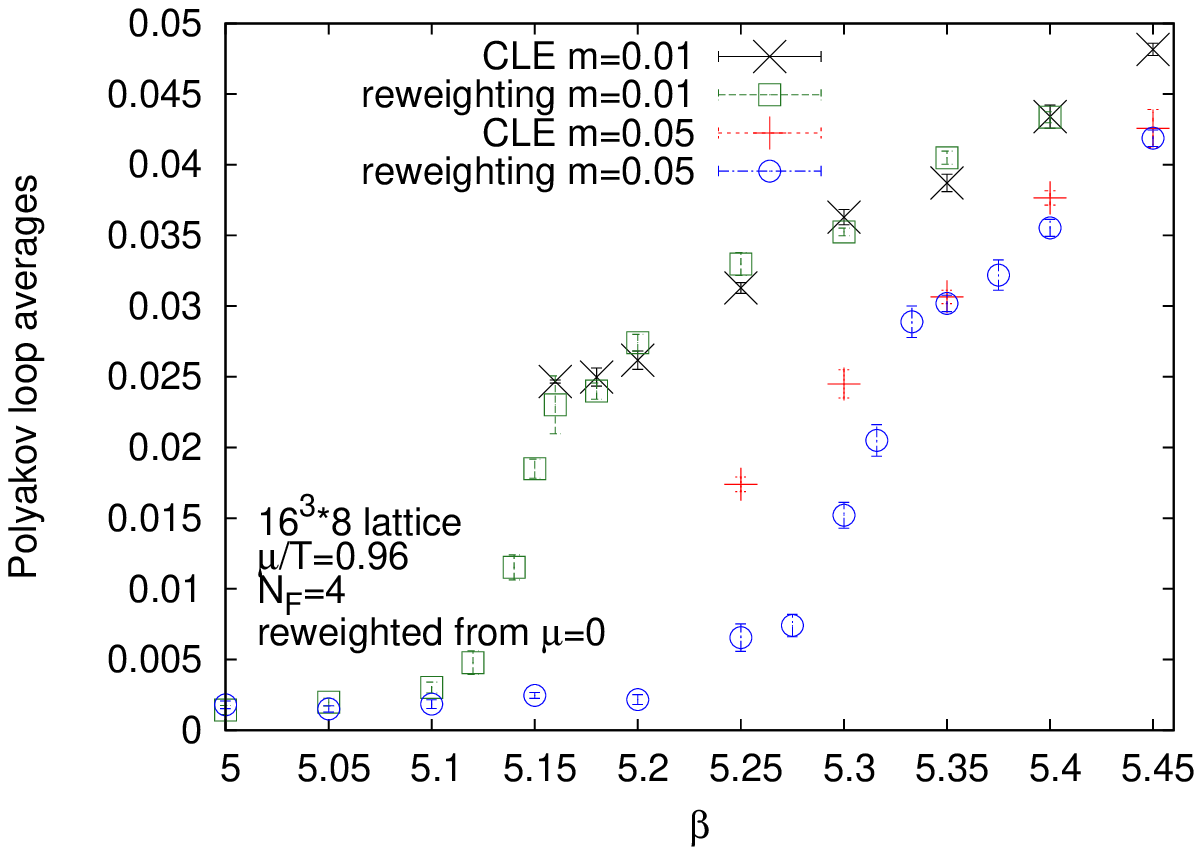, width=8cm}
 \caption{Comparison of the plaquette averages and Polyakov loops 
as a function of the 
$\beta$ parameter at a fixed $\mu/T=0.96$, using  $m=0.01$ 
and $m=0.05$ as indicated, 
calculated with CLE and reweighting from the $\mu=0$ ensemble on
$16^3*8$ lattices.
 }
\label{fig-nt8-vertical} 
\end{center}
\end{figure}

\begin{figure}
\begin{center}
 \epsfig{file=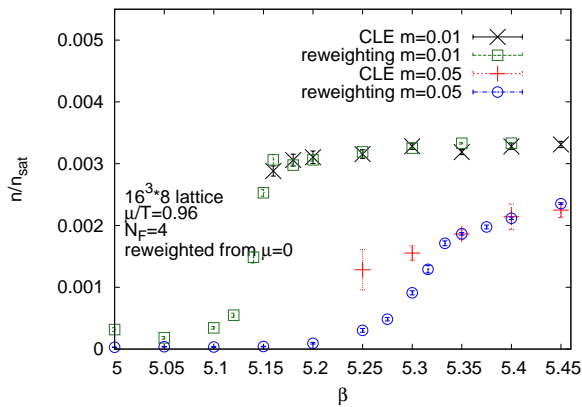, width=8cm}
 \caption{Comparison of the fermionic density  
as a function of the 
$\beta$ parameter at a fixed $\mu/T=0.96$, using $m=0.01$ 
and $m=0.05$ as indicated,
calculated with CLE and reweighting from the $\mu=0$ ensemble on
$16^3*8$ lattices.
 }
\label{fig-nt8-vertical-ferm} 
\end{center}
\end{figure}

 Finally we investigated $N_t=8$ lattices. In Fig.~\ref{fig-nt8-vertical} we 
show the behavior of the gauge observables, 
in Fig.~\ref{fig-nt8-vertical-ferm} the fermionic density. We used 
200-300 independent configurations at each $\beta$ value to 
perform the reweighting. At small betas the 
complex Langevin simulations become instable also on these lattices, 
which can be observed in Figs.~\ref{fig-nt8-vertical} 
and~\ref{fig-nt8-vertical-ferm} by the absence of results.
One observes that the CLE breaks down above the lattice spacing 
$ a \approx 0.15 \textrm{fm} $.

\section{Conclusions}
\label{concsec}

In this paper we have compared complex Langevin 
simulations of finite density QCD with 
reweighting from the positive ensembles of the phasequenched theory and $\mu=0$.

Both methods have a limited region of parameter space where they are 
applicable. The complex Langevin method fails for too small $\beta$ 
parameters, as noted earlier, 
but this still allows the exploration of the whole phase diagram in 
HDQCD \cite{Aarts:2014kja}. Reweighting from zero $\mu$ breaks down because of the overlap and sign problems around $ { \mu\over T} \approx 1 -1.5 $. 
In contrast, the reweighting from the phasequenched ensemble in the 
deconfined phase performs better also for large $\mu$, suggesting 
that the sign problem is not that severe.

We observe good agreement of these two methods in the region where they 
are both applicable. The failure of both methods can be assessed independently 
of the comparison: the complex Langevin simulations
develop 'skirted' distributions as the gauge cooling loses its effectiveness,
and the errors of the reweighting start to grow large signaling 
sign and overlap problems.

An important question for the applicability of the complex Langevin method 
to explore the phase diagram of QCD is the behavior of $\beta_{min}$, 
the lattice parameter below which gauge cooling is not effective. In 
this study we have determined that using $N_t= 4, N_t=6 $ 
and $N_t=8$ lattices (with pion mass $ m_\pi/T_c \approx 2.2 -2.4 $)
this breakdown prevents the exploration of the 
deconfinement transition and the location of a possible critical point.

\acknowledgments \noindent
This project was funded by the DFG grant SFB/TR55. S. D. Katz is funded by
the "Lend\"ulet" program of the Hungarian Academy of Sciences
((LP2012-44/2012).
The authors gratefully acknowledge the Gauss Centre for Supercomputing 
e.V. (www.gauss-centre.eu) for funding this project by providing computing 
time on the GCS Supercomputer SuperMUC at Leibniz Supercomputing 
Centre (LRZ, www.lrz.de). Some parts of the numerical calculation were 
 done on the GPU cluster at Eotvos and Wuppertal Universities.

\bibliography{mybib}
  
\end{document}